\documentclass[aps,prd,twocolumn,groupedaddress,showpacs,nofootinbib]{revtex4}
\usepackage{graphicx}

\begin{document}

\title{$f(R)$ theories of gravity in Palatini
approach matched with observations}

\author{S. Capozziello, V.F.
Cardone }\affiliation{Dipartimento di Fisica ``E.R. Caianiello'',
Universit\`{a} di Salerno and INFN, Sez. di Napoli, Gruppo Coll.
di Salerno, via S. Allende, 84081 - Baronissi (Salerno), Italy}
\author{M. Francaviglia}
\affiliation{Dipartimento di Matematica, Universit\`{a} di Torino,
Via C. Alberto 10, 10123 - Torino, Italy}
\email{capozziello@sa.infn.it,  winny@na.infn.it,
francaviglia@dm.unito.it}
\begin{abstract}

We investigate the viability of $f(R)$ theories in the framework
of the Palatini approach as solutions to the problem of the
observed accelerated expansion of the universe. Two physically
motivated popular choices for $f(R)$ are considered\,: power law,
$f(R) = \beta R^n$, and logarithmic, $f(R) = \alpha \ln{R}$. Under
the Palatini approach, both Lagrangians give rise to cosmological
models comprising only standard matter and undergoing a present
phase of accelerated expansion. We use the Hubble diagram of type
Ia Supernovae and the data on the gas mass fraction in relaxed
galaxy clusters to see whether these models are able to reproduce
what is observed and to constrain their parameters. It turns out
that they are indeed able to fit the data with values of the
Hubble constant and of the matter density parameter in agreement
with some model independent estimates, but the today deceleration
parameter is higher than what is measured in the concordance
$\Lambda$CDM model.

\end{abstract}


\maketitle

\section{Introduction}

The Hubble diagram of type Ia supernovae (hereafter SNeIa)
\cite{SNeIa,Riess04}, the anisotropy spectrum of the cosmic
microwave background radiation (hereafter CMBR)
\cite{CMBR,WMAP,VSA}, the matter power spectrum determined by the
large scale distribution of galaxies \cite{LSS,Teg03} and by the
data on the Ly$\alpha$ clouds \cite{Lyalpha} are all convincing
evidences in favour of a new picture of the universe, unexpected
only few years ago. According to this nowadays standard scenario,
the universe is flat and undergoing an accelerated expansion
driven by a mysterious fluid with negative pressure nearly
homogeneously distributed and making up to $\sim 70\%$ of the
energy content. This exotic component is what is called {\it dark
energy}, while the model we have just depicted is usually referred
to as the {\it concordance model}.

Even if strongly supported by the bulk of the available
astrophysical data, this new picture is not free of problems.
Actually, while it is clear how dark energy works, its nature
remains an unsolved problem. The simplest explanation claims for
the cosmological constant $\Lambda$ thus leading to the so called
$\Lambda$CDM model\footnote{It is common in literature to make no
distinction between the concordance and the $\Lambda$CDM model
even if, strictly speaking, in the concordance model the dark
energy may also be provided by a different mechanism.}
\cite{Lambda}. Although being the best fit to most of the
available astrophysical data \cite{WMAP,Teg03}, the $\Lambda$CDM
model is also plagued by many problems on different scales. If
interpreted as vacuum energy, $\Lambda$ is up to 120 orders of
magnitudes smaller than the predicted value. Furthermore, one
should also solve the {\it coincidence problem}, i.e. the nearly
equivalence of the matter and $\Lambda$ contribution to the total
energy density. As a response to these problems, much interest has
been devoted to models with dynamical vacuum energy, dubbed {\it
quintessence} \cite{QuintFirst}. These models typically involve a
scalar field rolling down its self interaction potential thus
allowing the vacuum energy to become dominant only recently (see
\cite{PR02,Pad02} for good reviews). Although quintessence by a
scalar field is the most studied candidate for dark energy, it
generally does not avoid {\it ad hoc} fine tuning to solve the
coincidence problem. Moreover, it is not clear where this scalar
field comes from and how to choose the self interaction potential.

On the other hand, it is worth noting that, despite the broad
interest in dark matter and dark energy, their physical properties
are still poorly understood at a fundamental level and, indeed, it
has never been shown that they are in fact two different
ingredients. This observation motivated the great interest
recently devoted to a completely different approach to
quintessence. Rather than fine tuning a scalar field potential, it
is also possible to explain the acceleration of the universe by
introducing a cosmic fluid with an exotic equation of state
causing it to act like dark matter at high density and dark energy
at low density. An attractive feature of these models is that they
can explain both dark energy and dark matter with a single
component (thus automatically solving the coincidence problem) and
have therefore been referred to as {\it unified dark energy} (UDE)
or {\it unified dark matter} (UDM).  Some interesting examples are
the generalized Chaplygin gas \cite{Chaplygin}, the tachyonic
field \cite{tachyon}, the condensate cosmology \cite{Bruce} and
the Hobbit model \cite{Hobbit}. It is worth noting, however, that
such models are seriously affected by problems with structure
formation \cite{Sandvik} so that some work is still needed before
they can be considered as reliable alternatives to dark energy.

Actually, there is still a different way to face the problem of
cosmic acceleration. As  stressed in Lue et al. \cite{LSS03}, it
is possible that the observed acceleration is not the
manifestation of another ingredient in the cosmic pie, but rather
the first signal of a breakdown of our understanding of the laws
of gravitation. From this point of view, it is thus tempting to
modify the Friedmann equations to see whether it is possible to
fit the astrophysical data with a model comprising only the
standard matter. Interesting examples of this kind are the
Cardassian expansion \cite{Cardassian} and the DGP gravity
\cite{DGP}.

In this same framework, there is also the attractive possibility
to consider the Einsteinian general relativity as a particular
case of a more fundamental theory. This is the underlying
philosophy of what are referred to as $f(R)$ theories
\cite{capozcurv,MetricRn,review,PalRn,lnR,ABF04}. In this case,
the Friedmann equations have to be given away in favour of a
modified set of cosmological equations that are obtained by
varying a generalized gravity Lagrangian where the scalar
curvature $R$ has been replaced by a generic function $f(R)$. The
usual general relativity is recovered in the limit $f(R) = R$,
while completely different results may be obtained for other
choices of $f(R)$. While in the weak field limit the theory should
give the usual newtonian gravity, at cosmological scales there is
an almost complete freedom in the choice of $f(R)$ thus leaving
open the way to a wide range of models.

The key point of $f(R)$ theories is the presence of modified
Friedmann equations obtained by varying the generalized
Lagrangian. However, here lies also the main problem of this
approach since it is not clear how the variation has to be
performed. Actually, once the Robertson\,-\,Walker metric has been
assumed, the equations governing the dynamics of the universe are
different depending on whether one varies with respect to the
metric only or with respect to the metric components and the
connections. It is usual to refer to these two possibilities as
the {\it metric} and the {\it Palatini approach} respectively. The
two methods give the same result only in the case $f(R) = R$,
while lead to significantly different dynamical equations for
every other choice of $f(R)$ (see \cite{MFF1987,FFV,ABF04,ACCF}
and references therein).

It is worth noting $f(R)$ theories were initially investigated
using the metric approach \cite{capozcurv,MetricRn,review}. Even
if some interesting and successful results have been obtained
\cite{curvfit}, this way to $f(R)$ theories is plagued by serious
mathematical difficulties. Actually, even for the simplest $f(R)$,
the metric approach leads to a fourth order nonlinear differential
equation for the scale factor that is impossible to solve
analytically and is affected by several problems that greatly
complicate the search for numerical solutions. Moreover, some
doubts have been cast on the consistency among the weak field
limit of the theory and the newtonian gravity as tested at the
Solar system scale \cite{newtlimitno} even if some interesting
different results have also been obtained \cite{newtlimitok}.

On the other hand, theoretical considerations about the stability
of the equations and the newtonian limit argue in favor of the
Palatini approach to $f(R)$ theories. Moreover, the dynamics of
the universe may be analytically determined from the cosmological
equations obtained with this method for some interesting cases. To
this aim, a clear mathematical machinery has been presented in
Ref.\,\cite{ABF04} (hereafter ABF04) that allows to determine
analytic expressions for the Hubble parameter as function of the
redshift. As we will see later, this is all what is needed to test
a given cosmological model.

The Palatini approach to $f(R)$ theories has been widely studied
in literature \cite{PalRn,lnR,MFF1987,FFV,ABF04,ACCF} and the
dynamics of the cosmological models obtained by applying this
method to different choices of $f(R)$ has been investigated in
detail. Here we adopt an observational point of view on the
Palatini approach. Assuming that this is the correct way to treat
$f(R)$ theories, we investigate the viability of two classes of
models obtained by two popular choices for $f(R)$, namely the
power law $f(R) = \beta R^n$ and the logarithmic $f(R) = \alpha
\ln{R}$. To this aim, we compare the model predictions against the
SNeIa Hubble diagram and the data on the gas mass fraction in
relaxed galaxy clusters. This analysis will allow us to constrain
the model parameters and to see whether $f(R)$ theories are indeed
reliable alternatives to dark energy. Moreover, this will be an
observational validation of the theoretically motivated Palatini
approach.

The paper is organized as follows. Sect.\,II details the method we
employ to constrain the models and present the dataset we will
use. The two classes of models we consider are briefly discussed
in Sect.\,III where we also individuate the parameters that are
better suited to both assign the model and be constrained by the
data. A detailed discussion of the results is the subject of
Sect.\,IV, while we summarize and conclude in Sect.\,V.

\section{Constraining a model}

Considered for a long time a purely theoretical science, cosmology
has today entered the realm of observations since it is now
possible to test cosmological predictions against a meaningful set
of astrophysical data. Much attention, in this sense, has been
devoted to standard candles, i.e. astrophysical objects whose
absolute magnitude $M$ is known (or may be exactly predicted) {\it
a priori} so that a measurement of its apparent magnitude $m$
immediately gives the distance modulus $\mu = m - M$. The distance
to the object is then estimated as\,:

\begin{equation}
\mu(z) = 5 \log{D_L(z)} + 25
\label{eq: distmod}
\end{equation}
with $D_L(z)$ the luminosity distance (in Mpc) and $z$ the
redshift of the object. The relation between $\mu$ and $z$ is what
is referred to as Hubble diagram and is an open window on the
cosmography of the universe. Furthermore, the Hubble diagram is a
powerful cosmological test since the luminosity distance is
determined by the expansion rate as\,:

\begin{equation}
D_L(z) = \frac{c}{H_0} (1 + z) \int_{0}^{z}{\frac{d\zeta}{E(\zeta)}}
\label{eq: dl}
\end{equation}
with $E(z) = H(z)/H_0$, $H = \dot{a}/a$ the Hubble parameter and
$a(t)$ the scale factor. Note that an overdot means
differentiation with respect to cosmic time, while an underscript
$0$ denotes the present day value of a quantity.

Being the Hubble diagram related to the luminosity distance and
being $D_L$ determined by the expansion rate $H(z)$, it is clear
why it may be used as an efficient tool to test cosmological
models and constrain their parameters. To this aim, however, it is
mandatory that the relation $\mu = \mu(z)$ is measured up to high
enough redshift since, for low $z$, $D_L$ reduces to a linear
function of the redshift (thus recovering the Hubble law) whatever
the background cosmological model is. This necessity claims for
standard candles that are bright enough to be visible at such high
redshift that the Hubble diagram may discriminate among different
rival theories. SNeIa are, up to now, the objects that best match
these requirements. It is thus not surprising that the first
evidences of an accelerating universe came from the SNeIa Hubble
diagram \cite{SNeIa} and dedicated survey (like the SNAP satellite
\cite{SNAP}) have been planned in order to increase the number of
SNeIa observed and the redshift range probed.

The most updated and reliable compilation of SNeIa is the {\it
Gold} dataset recently relased by Riess et al. \cite{Riess04}. The
authors have compiled a catalog containing 157 SNeIa with $z$ in
the range $(0.01, 1.70)$ and visual absorption $A_V < 0.5$. The
distance modulus of each object has been evaluated by using a set
of calibrated methods so that the sample is homogenous in the
sense that all the SNeIa have been re-analyzed using the same
technique in such a way that the resulting Hubble diagram is
indeed reliable and accurate. Given a cosmological model assigned
by a set of parameters ${\bf p} = (p_1, \ldots, p_n)$, the
luminosity distance may be evaluated with Eq.(\ref{eq: dl}) and
the predicted Hubble diagram contrasted with the observed SNeIa
one. Constraints on the model parameters may then be extracted by
mean of a $\chi^2$\,-\,based analysis defining the $\chi^2$ as\,:

\begin{equation}
\chi_{SNeIa}^2 = \sum_{i = 1}^{N_{SNeIa}}{\left [ \frac{\mu(z_i, {\bf p}) - \mu_{obs}(z_i)}{\sigma_i} \right ]^2}
\label{eq: chisneia}
\end{equation}
where $\sigma_i$ is the error on the distance modulus at redshift
$z_i$ and the sum is over the $N_{SNeIa}$ SNeIa observed. It is
worth stressing that the uncertainty on each measurement also
takes into account the error on the redshifit and are not gaussian
distributed. As a consequence, the reduced $\chi^2$ (i.e.,
$\chi_{SNeIa}^2$ divided by the number of degrees of freedom) for
the best fit model is not forced to be close to unity.
Nonetheless, different models may still be compared on the basis
of the $\chi^2$ value\,: the lower is $\chi_{SNeIa}^2$, the better
the model fits the SNeIa Hubble diagram.

The method outlined above is a simple and quite efficient way to
test whether a given model is a viable candidate to describe the
late time evolution of the universe. Nonetheless, it is affected
by some degeneracies that could be only partially broken by
increasing the sample size and extending the redshift range
probed. A straightforward example may help in elucidating this
point. Let us consider the flat concordance cosmological model
with matter and cosmological constant. It is\,:

\begin{displaymath}
E^2(z) = \Omega_M (1+z)^3 + (1 - \Omega_M)
\end{displaymath}
so that $\chi_{SNeIa}^2$ will only depend on the Hubble constant
$H_0$ and the matter density parameter $\Omega_M$. Actually, we
could split the matter term in a baryonic and a non baryonic part
denoting with $\Omega_b$ the baryon density parameter. Since both
baryons and non baryonic dark matter scales as $(1 + z)^3$, $E(z)$
and thus the luminosity distance will depend only on the total
matter density parameter and we could never constrain $\Omega_b$
by fitting the SNeIa Hubble diagram. Similar degeneracies could
also happen with other cosmological models thus stressing the need
for complementary probes to be combined with the SNeIa data.

To this aim, we consider a recently proposed test based on the gas
mass fraction in galaxy clusters. We briefly outline here the
method referring the interested reader to the literature for
further details \cite{fgasbib,fgasapp}. Both theoretical arguments
and numerical simulations predict that the baryonic mass fraction
in the largest relaxed galaxy clusters should be invariant with
the redshift (see, e.g., Ref.\,\cite{ENF98}). However, this will
only appear to be the case when the reference cosmology in making
the baryonic mass fraction measurements matches the true
underlying cosmology. From the observational point of view, it is
worth noting that the baryonic content in galaxy clusters is
dominated by the hot X\,-\,ray emitting intra-cluster gas so that
what is actually measured is the gas mass fraction $f_{gas}$ and
it is this quantity that should be invariant with the redshift
within the caveat quoted above. Moreover, it is expected that the
baryonic fraction in clusters equals the universal ratio
$\Omega_b/\Omega_M$ so that $f_{gas}$ should indeed be given by $b
{\times} \Omega_b/\Omega_M$ where the multiplicative factor $b$ is
motivated by simulations that suggest that the gas fraction is
slightly lower than the universal ratio because of processes that
convert part of the gas into stars or eject it outside the cluster
itself.

Following Ref.\,\cite{fgasdata} (hereafter A04), we adopt the SCDM
model (i.e., a flat universe with $\Omega_M = 1$ and $h = 0.5$,
being $h$ the Hubble constant in units of $100 \ {\rm km \ s^{-1}
\ Mpc^{-1}}$) as reference cosmology in making the measurements so
that the theoretical expectation for the apparent variation of
$f_{gas}$ with the redshift is \cite{fgasdata}\,:

\begin{equation}
f_{gas}(z) = \frac{b \Omega_b}{(1 + 0.19 \sqrt{h}) \Omega_M} \left [ \frac{D_A^{SCDM}(z)}{D_A^{mod}(z)} \right ]^{1.5}
\label{eq: fgas}
\end{equation}
where $D_A^{SCDM}$ and $D_A^{mod}$ is the angular diameter
distance for the SCDM and the model to be tested respectively.
$D_A(z)$ may be evaluated from the luminosity distance $D_L(z)$
as\,:

\begin{equation}
D_A(z) = (1 + z)^{-2} D_L(z)
\label{eq: da}
\end{equation}
with $D_L(z)$ given by Eq.(\ref{eq: dl}) above.

A04 have extensively analyzed the set of simulations in
Ref.\,\cite{ENF98} to get $b = 0.824 {\pm} 0.089$. In our analysis
below, we will set $b = 0.824$  in order to not increase the
number of parameters to be constrained. Actually, we have checked
that, for values in the $1 \sigma$ range quoted above, the main
results are independent on $b$. It is worth noting that, while the
angular diameter distance depends on $E(z)$ and thus on $h$ and
$\Omega_M$, the prefactor in Eq.(\ref{eq: fgas}) makes $f_{gas}$
explicitly depending on $\Omega_b/\Omega_M$ so that a direct
estimate of $\Omega_b$ is (in principle) possible. Actually, we
will see later that, for the models we will consider, the quantity
that is constrained by the data is the ratio $\Omega_b/\Omega_M$
rather than $\Omega_b$ itself.

To simultaneously take into account both the fit to the SNeIa
Hubble diagram and the test on the $f_{gas}$ data, it is
convenient to perform a likelihood analysis defining the following
likelihood function\,:

\begin{equation}
{\cal{L}}({\bf p}) \propto \exp{\left [ - \frac{\chi^2({\bf p})}{2} \right ]}
\label{eq: deflike}
\end{equation}
with\,:

\begin{equation}
\chi^2 = \chi_{SNeIa}^2 + \chi_{gas}^2 + \left ( \frac{h - 0.72}{0.08} \right )^2 + \left ( \frac{\Omega_b/\Omega_M - 0.16}{0.06} \right )^2
\label{eq: defchi}
\end{equation}
where we have defined\,:

\begin{equation}
\chi_{gas}^2 =
\sum_{i = 1}^{N_{gas}}{\left [ \frac{f_{gas}(z_i, {\bf p}) - f_{gas}^{obs}(z_i)}{\sigma_{gi}} \right ]^2}
\label{eq: chigas}
\end{equation}
being $f_{gas}^{obs}(z_i)$ the measured gas fraction in a galaxy
clusters at redshift $z_i$ with an error $\sigma_{gi}$ and the sum
is over the $N_{gas}$ clusters considered. In order to avoid
possible systematic errors in the $f_{gas}$ measurement, it is
desirable that the cluster is both highly luminous (so that the
S/N ratio is high) and relaxed so that both merging processes and
cooling flows are absent. A04 \cite{fgasdata} have recently
released a catalog comprising 26 large relaxed clusters with a
precise measurement of both the gas mass fraction $f_{gas}$ and
the redshift $z$ (not presented in the quoted paper). We use these
data to perform our likelihood analysis in the following.

Note that, in Eq.(\ref{eq: defchi}), we have explicitly introduced
two gaussian priors to better constrain the model parameters.
First, there is a prior on the Hubble constant $h$ determined by
the results of the HST Key project \cite{HSTKey} from an accurate
calibration of a set of different local distance estimators.
Further, we impose a constraint on the ratio $\Omega_b/\Omega_M$
by considering the estimates of $\Omega_b h^2$ and $\Omega_M h^2$
obtained by Tegmark et al. \cite{Teg03} from a combined fit to the
SNeIa Hubble diagram, the CMBR anisotropy spectrum measured by
WMAP and the matter power spectrum extracted from over 200000
galaxies observed by the SDSS collaboration. It is worth noting
that, while our prior on $h$ is the same as that used by many
authors when applying the $f_{gas}$ test \cite{fgasapp,fgasdata},
it is common to put a second prior on $\Omega_b$ rather than
$\Omega_b/\Omega_M$. Actually, this choice is motivated by the
peculiar features of the models we will consider that lead us to
choose this unusual prior for reasons that will be clear later.

With the definition (\ref{eq: deflike}) of the likelihood
function, the best fit model parameters are those that maximize
${\cal{L}}({\bf p})$. However, to constrain a given parameter
$p_i$, one resorts to the marginalized likelihood function defined
as\,:

\begin{equation}
{\cal{L}}_{p_i}(p_i) \propto \int{dp_1 \ldots \int{dp_{i - 1} \int{dp_{i + 1} \ldots \int{dp_n {\cal{L}}({\bf p})}}}}
\label{eq: defmarglike}
\end{equation}
that is normalized at unity at maximum. The $1 \sigma$ confidence
regions are determined by $\delta \chi^2 = \chi^2 - \chi_0^2 = 1$,
while the condition $\delta \chi^2  = 4$ delimited the $2 \sigma$
confidence regions. Here, $\chi_0^2$ is the value of the $\chi^2$
for the best fit model. Projections of the likelihood function
allow to show eventual correlations among the model parameters. In
these two dimensional plots, the $1 \sigma$ and $2 \sigma$ regions
are formally defined by $\Delta \chi^2 = 2.30$ and $6.17$
respectively so that these contours are not necessarily equivalent
to the same confidence level for single parameter estimates.

\section{The $f(R)$ models}

The observed cosmic acceleration is currently explained by
invoking the presence of a new fluid with negative pressure which
smoothly fills the universe driving its expansion. However, the
nature and the nurture of this fluid are yet unknown so that other
radically different proposals, such as unified dark energy models
\cite{Chaplygin,tachyon,Bruce,Hobbit} or brane world inspired
theories \cite{LSS03,DGP}, are still viable and worth exploring.

A quite interesting and fascinating scenario predicts that
standard matter is the only ingredient of the cosmic pie as it is
indeed observed, but the Einsteinian general relativity breaks
down at the present small curvature scale. As a result, one should
generalize the action as\,:

\begin{displaymath}
{\cal{A}} = \int{\left [ \sqrt{g} f(R) + 2 \kappa L_{mat} \right ] d^4x}
\end{displaymath}
with $\kappa = 8 \pi G$ and $L_{mat}$ the matter Lagrangian.
Varying with respect to the metric components and adopting then
the Robertson\,-\,Walker metric, one obtains modified Friedmann
equations that, by rearranging suitably the different terms, may
still be formally written in the same way as the usual ones
provided that a new fictitious component is added. For instance,
the Hubble parameter is now given as\,:

\begin{equation}
H^2 = \frac{\kappa}{3} \left ( \rho_m + \rho_{curv} \right )
\label{eq: hubblecurv}
\end{equation}
with $\rho_m$ the standard matter energy density and $\rho_{curv}$
the energy density of a {\it curvature fluid} whose density and
pressure are given in terms of $f(R)$ and its derivatives (see
\cite{capozcurv,review} for details). Although intriguing, this
approach leads to a mathematically untractable problem. Indeed, it
turns out that the scale factor $a(t)$ should be obtained by
solving a nonlinear fourth order differential equation. Not
surprisingly, it is not possible to analytically solve this
equation even for the simplest choices of $f(R)$.  Moreover, some
conceptual difficulties make it a daunting task to look for
numerical solutions.

An attractive way to escape these problems is to resort to the so
called {\it Palatini approach} in which the field equations are
obtained by varying with respect to both the metric components and
the connections considered as independent variables. A consistency
condition is then imposed to complement the system thus giving a
set of first order differential equations for the scale factor
$a(t)$ and the scalar curvature $R$. The modified Friedmann
equations are finally obtained by imposing that the metric is the
Robertson\,-\,Walker one (see, e.g., \cite{ABF04} for a clear
illustration of the procedure).

The Palatini approach is physically well motivated and has the
attractive feature that the Hubble parameter $H(z)$, that is all
what is needed for constraining the model, may be expressed
analytically for some choices of the function $f(R)$. It is thus
quite interesting to constrain the cosmological models obtained by
applying the Palatini approach with two different choices of the
function $f(R)$. The main characteristics of these models are
briefly presented below. We follow Ref.\,\cite{ABF04} (hereafter
ABF04) which the interested reader is referred to for further
details.

\subsection{The Power law Lagrangian}

We first consider the class of Lagrangians that are linear in an
aribtrary power of the scalar curvature $R$\,:

\begin{equation}
f(R) = \beta R^n
\label{eq: plfr}
\end{equation}
with $\beta \ne 0$ and $n \ne 0,2$ real parameters to be
constrained. Note that $\beta$ has the same units of $R^{n}$ so
that $f(R)$ is adimensional. This model has been already discussed
by many authors \cite{capozcurv,MetricRn,review} using the
standard way of varying the Lagrangian. In particular, in
Ref.\,\cite{curvfit}, some of us have also successfully tested a
simplified version of this model (with no matter term) against the
SNeIa Hubble diagram. Moreover, this kind of Lagrangian has also
been investigated in the framework of the Palatini approach
\cite{PalRn,ABF04}. It is thus particularly interesting to see
whether the Palatini approach leads to results that are in
agreement with the observed data. Using the same notation as in
ABF04, the scale factor $a(t)$ and the Hubble parameter $H(z)$ for
a flat universe are given as\,:

\begin{equation}
a(t) = \left [ \frac{3 \epsilon}{2 n (3 - n)} \right ]^{n/3} \left [ \frac{\kappa \eta}{\beta (2 - n)} \right ]^{1/3} t^{2 n/3} \ ,
\label{eq: atpl}
\end{equation}

\begin{equation}
H^2(z) = \frac{2 \epsilon n (\kappa \eta)^{1/n}}{3 (3 - n) \left [ \beta (2 - n) (1 + z)^{-3} \right ]^{1/n}} \ ,
\label{eq: hzpl}
\end{equation}
with $\eta = \rho_m(z = 0)$ the present day value of the matter
density and $\epsilon = {\pm} 1$ depending on $n$ in such a way
that both $a(t)$ and $H(z)$ are correctly defined. For the
applications, it is better to use the following relation\,:

\begin{displaymath}
\kappa \eta = 3 \Omega_M H_0^2
\end{displaymath}
with $\Omega_M$ the usual matter density parameter. It is worth
stressing that, even if we assume a flat model, $\Omega_M$ is not
forced to be unity since the critical density for closure is now
different from the usual value $\rho_c = 3 H_0^2/8 \pi G$. The
present day age of the universe may be obtained by evaluating
Eq.(\ref{eq: atpl}) at the present day and then solving with
respect to $t_0$ thus obtaining\,:

\begin{equation}
t_0 = \left [ \frac{3 \epsilon}{2 n (3 - n)} \right ]^{-1/2} \left [ \frac{3 \Omega_M H_0^2}{\beta (2 - n)} \right ]^{- \frac{1}{2 n}} \ .
\label{eq: tzpl}
\end{equation}
Being the scale factor a power law function of the time, the
deceleration parameter is constant and given as\,:

\begin{equation}
q \equiv - \frac{a \ddot{a}}{\dot{a}^2} = \frac{3 - 2 n}{2 n}
\label{eq: qzpl}
\end{equation}
so that we may exclude all the Lagrangians with $n \le 3/2$ since
they give rise to non accelerating models $(q_0 \ge 0)$.

A nice feature of this model is that the dimensionless Hubble parameter is simply\,:

\begin{equation}
E^2(z) = (1 + z)^{3/n}
\label{eq: ezpl}
\end{equation}
so that the luminosity distance turns out to be\,:

\begin{equation}
D_L(z) = \frac{c}{H_0} \frac{2 n}{2 n - 3} (1 + z) \left [ (1 + z)^{\frac{2 n - 3}{2 n}} - 1 \right ] \ .
\label{eq: dlpl}
\end{equation}
Both $D_L$ and $D_A = (1 + z)^{-2} D_L$ depend only on the two
parameters $n$ and $H_0$ so that fitting to the SNeIa Hubble
diagram is unable to put any constraint neither on $\beta$ or
$\Omega_M$. Adding the test on the $f_{gas}$ data described in the
previous section partially alleviates this problem since
$f_{gas}(z)$ depends also on $\Omega_b/\Omega_M$. It is then
possible to get an estimate of $\Omega_M$ combining the constraint
on $\Omega_b/\Omega_M$ with an independent knowledge of $\Omega_b$
from the measured abundance of light elements or primordial
nucleosynthesis. Finally, the coupling parameter could be derived
inverting Eq.(\ref{eq: tzpl}) with respect to $\beta$ itself
provided that $t_0$ has been somehow evaluated (possibly from a
model independent method).

As a general remark, let us observe that, without a knowledge of
$t_0$, the parameter that can be constrained is $\Omega_M/\beta$.
Qualitatively, this could be explained by noting that all the
tests we are considering are related to the {\it cosmography} of
the universe. This is determined by the balance between the matter
content and the {\it exotic} geometrical effects due to the
replacement of $R$ with $f(R)$ in the gravity Lagrangian.
Actually, this feature is common to all $f(R)$ theories and could
be expected since now geometry plays the same role as the scalar
field in the usual dark energy models.

\subsection{The Logarithmic Lagrangian}

Quantum effects in curved spacetimes may induce logarithmic terms
in the gravity Lagrangian \cite{lnR}. It is thus interesting to
consider the choice\,:

\begin{equation}
f(R) = \alpha \ln{R}
\label{eq: logfr}
\end{equation}
where the dimensions of $\alpha$ are such that $f(R)$ is
dimensionless\footnote{Note that, in literature, it is sometimes
adopted the choice $f(R) = \alpha \ln{\beta_l R}$. We follow ABF04
and set $\beta_l = 1$ with no loss of generality.}. This model is
more complicated than the power law one so that, as a result, it
is not possible to derive an analytical expression for the scale
factor. However, the dimensionless Hubble parameter may still be
expressed analytically as\,:

\begin{eqnarray}
E^2(z) & = & \displaystyle{\left [ \frac{1 + (9/4) \Omega_M H_0^2 \alpha^{-1}}
{1 + (9/4) \Omega_M H_0^2 \alpha^{-1} (1 + z)^3} \right ]^2} \nonumber \\
~ & {\times} & \displaystyle{\frac{1 + 9 \Omega_M H_0^2 \alpha^{-1} (1 + z)^3}{1 + 9 \Omega_M H_0^2 \alpha^{-1}}} \nonumber \\
~ & {\times} & \exp{\left \{ (3/2) \Omega_M H_0^2 \alpha^{-1} \left [ (1 + z)^3 - 1 \right ] \right \}} \ .
\label{eq: hlog}
\end{eqnarray}
The luminosity density is obtained inserting Eq.(\ref{eq: hlog})
into the definition (\ref{eq: dl}). There is not an analytic
expression for $D_L$, but the integral is straightforward to
evaluate numerically for a given value of $\Omega_M H_0
\alpha^{-1}$. As a consequence, the likelihood function for this
model depend on the Hubble constant $H_0$, the ratio
$\Omega_b/\Omega_M$ between the baryonic and total matter density
and the combined parameter $\Omega_M H_0^2 \alpha^{-1}$. It is
worth stressing that, even if in principle possible, constraining
separately the three parameters $(\Omega_M, H_0, \alpha)$ is not
correct since both $D_L(z)$ and $f_{gas}(z)$ depend on $\alpha$
only through the combination $\Omega_M H_0^2 \alpha^{-1}$.
Henceforth, it is this quantity that is constrained by the data.
Actually, this degeneracy may be broken by an independent estimate
of $\Omega_b$ that can be combined with the constraint on
$\Omega_b/\Omega_M$ to evaluate $\Omega_M$ and then $\alpha$ from
the constrained $\Omega_M H_0^2 \alpha^{-1}$. Note that, without
an estimate of $\Omega_b$ the only quantities estimated from the
fit to the SNeIa Hubble diagram are $H_0$ and $\Omega_M H_0^2
\alpha^{-1}$ so that only the parameter $\Omega_m/\alpha$ may be
constrained as a result of the above mentioned degeneracy between
matter and geometry.

There is no explicit analytic expression for the age of the
universe so that one has to resort to numerical integration of the
following relation\,:

\begin{equation}
t_0 = 9.78 \ h^{-1} \int_{0}^{\infty}{\frac{d\zeta}{(1 + \zeta) H(\zeta)}}
\label{eq: tzlog}
\end{equation}
giving $t_0$ expressed in Gyr. Le us remark that, while for power
law Lagrangians $t_0$ and $\Omega_b$ are needed to break the
degeneracy $\Omega_M/\beta$, now $\Omega_b$ and the likelihood
analysis are sufficient to estimate both $\Omega_M$ and $\alpha$
so that $t_0$ may be used to check the results against an
independent quantity.

Another striking difference with the case of power law $f(R)$ is
the fact that the deceleration parameter is no longer constant.
Even if we do not have an analytic expression for $a(t)$, we may
still evaluate $q$ as follows\,:

\begin{displaymath}
q = -1 + \frac{1 + z}{H} \frac{dH}{dz} \ .
\end{displaymath}
Inserting Eq.(\ref{eq: hlog}) into the above relation and
evaluating the result at the present day ($z = 0$), we get\,:

\begin{eqnarray}
q_0 & = &  -1 + \displaystyle{\frac{\Omega_M H_0^2 \alpha^{-1}}{4}} \nonumber \\
~ & {\times} & \left ( 9 + \displaystyle{\frac{54}{1 + 9 \Omega_M H_0^2 \alpha^{-1}}} - \displaystyle{\frac{108}{4 + 9 \Omega_M H_0^2
\alpha^{-1}}} \right ) \ .
\label{eq: qlog}
\end{eqnarray}
Eq.(\ref{eq: qlog}) shows that $q_0$ depends only on the parameter
$\Omega_M H_0^2\alpha^{-1}$ that is therefore what determines
whether the universe is today accelerating or decelerating. It is
also worth noting that $q(z)$ (not explicitly reported here for
sake of shortness) changes sign during the evolution of the
universe so that it is possible to estimate a transition redshift
$z_T$ as $q(z_T) = 0$ that only depends on $\Omega_M H_0^2
\alpha^{-1}$. It should be possible to estimate somewhat $z_T$,
this could give an independent check of the results. Actually, we
will see that this is not possible since all the estimates of
$z_T$ are model dependent. However, it is interesting to compare
the transition redshift predicted for the logarithmic $f(R)$ with
that of other dark energy models.

\section{Results}

We have applied the method described in Sect.\,II to investigate
whether the cosmological models obtained by applying the Palatini
approach to $f(R)$ theories for the two choices in Eqs.(\ref{eq:
plfr}) and (\ref{eq: logfr}) are in agreement with both the SNeIa
Hubble diagram and the data on the gas mass fraction in relaxed
galaxy clusters. This also allows us to constrain the model
parameters and compare the estimated values of some of them (as
the Hubble constant $h$ and the matter density $\Omega_M$) with
the recent results in literature in order to see whether they are
reliable or not.

\subsection{$f(R) = \beta R^n$}

Let us first discuss the case of the power law Lagrangian. The
best fit parameters turn out to be\,:

\begin{figure}
\centering \resizebox{8.5cm}{!}{\includegraphics{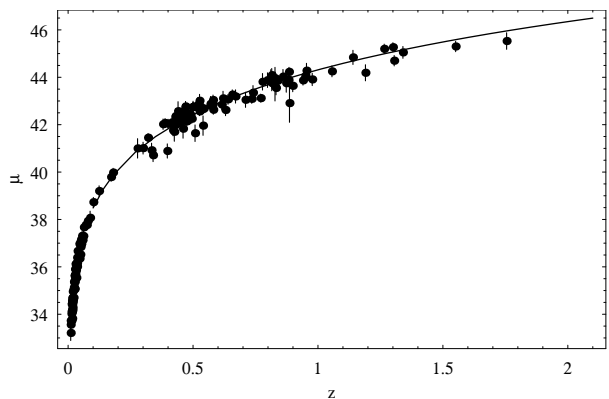}}
\caption{Best fit curve to the SNeIa Hubble diagram for the power
law Lagrangian model.} \label{fig: snefitpl}
\end{figure}

\begin{equation}
n = 2.25 \ \ , \ \ h = 0.641 \ \ , \ \ \Omega_b/\Omega_M = 0.181
\label{eq: bfpl}
\end{equation}
that gives the best fit curves shown in Figs.\,\ref{fig: snefitpl}
and \ref{fig: fgasfitpl}. The agreement with the data (in
particular, with the SNeIa Hubble diagram) is quite good which
should be considered a strong evidence in favor of the model.
However, Fig.\,\ref{fig: snefitpl} shows that the model slightly
overpredicts the distance modulus for two highest redshift
$SNeIa$, but, given the paucity of the data in this redshift
range, the discrepancy is hardly significative. Should this trend
be confirmed by future data (observable, e.g., with the SNAP
satellite mission that will detect SNeIa up to $z \sim 2$), we
should exclude the choice (\ref{eq: plfr}) for $f(R)$. Actually,
such a result could be expected since the deceleration parameter
is constant, while Riess et al. \cite{Riess04} claimed to have
detected a change in the sign of $q$ at a transition redshift $z_T
\sim 0.5$. We will return later to the problems connected with the
result of Riess et al. that lead us to consider (at least)
premature to deem as unreliable a model with a constant $q$.
Therefore, we still retain $f(R)$ theories with power law
Lagrangian.

\begin{figure}
\centering \resizebox{8.5cm}{!}{\includegraphics{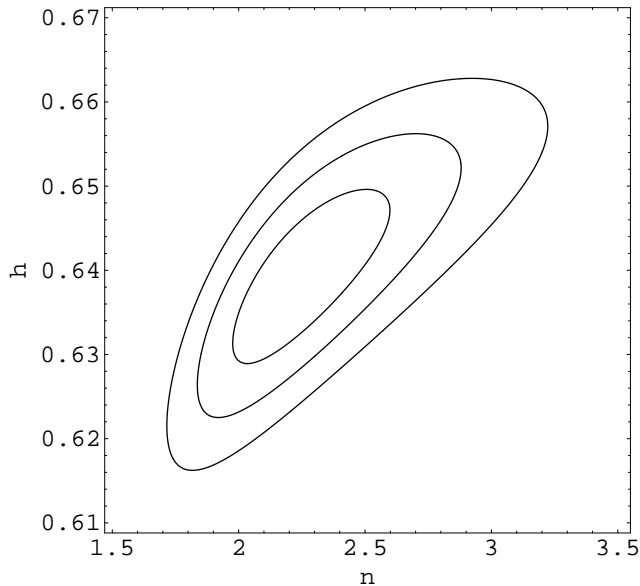}}
\caption{$1, \ 2, \ {\rm and} \ 3 \sigma$ confidence regions in the two dimensional parameter space $(n, h)$.}
\label{fig: nhrn}
\end{figure}

\begin{figure}
\centering \resizebox{8.5cm}{!}{\includegraphics{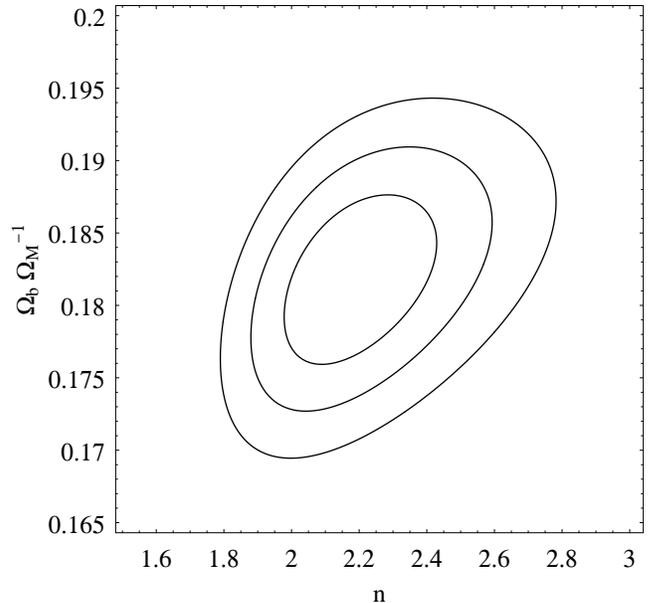}}
\caption{$1, \ 2, \ {\rm and} \ 3 \sigma$ confidence regions in the two dimensional parameter space $(n, \Omega_b/\Omega_M)$.}
\label{fig: nobomrn}
\end{figure}

It is interesting to look at the confidence contours in the
projected two parameters space. Figs.\,\ref{fig: nhrn} and
\ref{fig: nobomrn} show the confidence regions for the parameters
$(n, h)$ and $(n, \Omega_b/\Omega_M)$ respectively. It turns out
that $n$ is positively correlated with both $h$ and
$\Omega_b/\Omega_M$ so that the higher is $n$, the higher is the
expansion rate and the lower is the matter content $\Omega_M$. As
a consequence, to fit the available data, models with steeper
(higher $n$) power law Lagrangians should contain less matter
which is a result disfavoring values of $n$ much larger than our
best fit.

Using the method detailed in Sect.\,II, we have obtained the
following constraint on the model parameters\,:

\begin{figure}
\centering \resizebox{8.5cm}{!}{\includegraphics{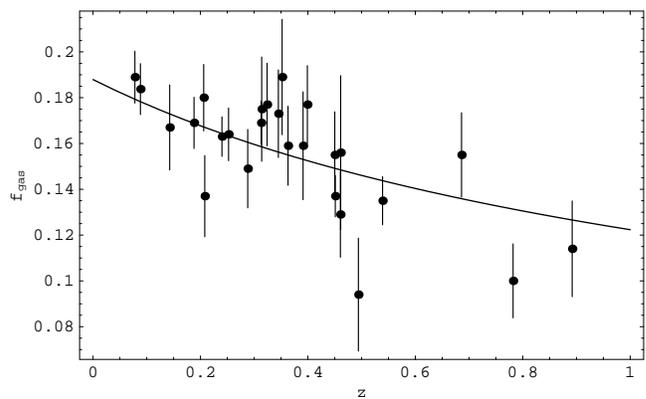}}
\caption{Best fit curve to the $f_{gas}$ data for the power law
Lagrangian model.} \label{fig: fgasfitpl}
\end{figure}

\begin{equation}
n \in \left \{
\begin{array}{ll}
(2.06, 2.46) & {\rm at} \ 1 \sigma \\ ~ & ~ ~ ~ \ \ \ \ \ \ \ ; \\ (1.91,
2.61) & {\rm at} \ 2 \sigma
\end{array}
\right .
\label{eq: nrpl}
\end{equation}

\begin{equation}
h \in \left \{
\begin{array}{ll}
(0.637, 0.648) & {\rm at} \ 1 \sigma \\ ~ & ~ ~ ~ \ \ \ \ \ \ \ ; \\
(0.633, 0.654) & {\rm at} \ 2 \sigma
\end{array}
\right .
\label{eq: hrpl}
\end{equation}

\begin{equation}
\Omega_b/\Omega_M \in \left \{
\begin{array}{ll}
(0.177, 0.185) & {\rm at} \ 1 \sigma \\ ~ & ~ ~ ~ \ \ \ \ \ \ \ . \\
(0.173, 0.189) & {\rm at} \ 2 \sigma
\end{array}
\right .
\label{eq: obompl}
\end{equation}
The cosmological model originating from power law $f(R)$ has been
already considered by different authors in literature under the
metric approach to the variation of the Lagrangian
\cite{capozcurv,MetricRn,review}. However, the lack of analytic
solutions for the scale factor or the Hubble parameter has
prevented any attempt to constrain the value of $n$ against the
observed data. Actually, only the model without matter has been
investigated giving $n \in (-0.450, -0.370)$ or $n \in (1.366,
1.376)$ \cite{curvfit} in clear disagreement with our estimate
(\ref{eq: nrpl}). However, such a comparison is meaningless
because of the presence of the matter term in the present case and
the absence in the other one.

Actually, using Eq.(\ref{eq: qzpl}), it is possible to convert the estimate of $n$ in a constraint on the present day value of the deceleration parameter. The best fit value for $n$ thus translates into $q_0 = -0.33$, while, combining Eqs.(\ref{eq: qzpl}) and (\ref{eq: nrpl}), we get\,:

\begin{equation}
q_0 \in \left \{
\begin{array}{ll}
(-0.39, -0.27) & {\rm at} \ 1 \sigma \\ ~ & ~ ~ ~ \ \ \ \ \ \ \ . \\
(-0.43, -0.21) & {\rm at} \ 2 \sigma
\end{array}
\right .
\label{eq: qzrpl}
\end{equation}
While consistent with the picture of an accelerating universe, our
estimates for $q_0$ disagree with other recent results. Let us
consider what is obtained for the $\Lambda$CDM model\footnote{We
limit our attention to the $\Lambda$CDM model only since the
cosmological constant is the simplest and most efficient way to
fit  most of the astrophysical data \cite{Teg03}. Moreover, the
constraints on the equation of state parameter $w = p/\rho$ are
still consistent with the cosmological constant value $w = -1$
\cite{EstW}. This conclusion is further strenghtened by the
methods that aim at recovering the evolution of the dark energy
density from the data in model independent way (see, e.g.,
\cite{WT04} and references therein).}. Using a flat geometry prior
and fitting to the SNeIa Hubble diagram only, Riess et al.
\cite{Riess04} have found $\Omega_M = 0.29^{+0.05}_{-0.03}$ that
gives\footnote{Hereafter, we will compute the error on $q_0$
propagating the maximum $1 \sigma$ uncertainty on $\Omega_M$.
Although not statistically correct, this method gives a quick
order of magnitude estimate of the error which is enough for our
aims.} $q_0 = -0.56 {\pm} 0.07$ that is not consistent with our
estimate. Adding the data on the CMBR anisotropy and the power
spectrum of SDSS galaxies, Tegmark et al. \cite{Teg03} give
$\Omega_M = 0.30 {\pm} 0.04$ so that the estimated $q_0$ is in
agreement with Riess et al. and hence in contrast with our value.
A similar result has also been obtained by A04 only using the same
$f_{gas}$ data we have considered here with a prior on $h$ and
$\Omega_b h^2$. For a flat $\Lambda$CDM model, their analysis
gives $\Omega_M = 0.24 {\pm} 0.04$ and hence $q_0 = -0.64 {\pm}
0.06$ still in disagreement with our Eq.(\ref{eq: qzrpl}). As a
general remark, we notice that our models turn out to be {\it less
accelerating} (i.e., the predicted $q_0$ is higher) than is
observed for the standard concordance model. From a different
point of view, lower values of $q_0$ correspond to higher $n$,
i.e. to steeper power law Lagrangians that are, however,
disfavoured by the lower matter content of the corresponding best
fit model.

However, one could deem as unreliable the comparison among $q_0$
constraints obtained under different underlying cosmological
models and look for model independent estimates of the
deceleration parameter. For instance, Riess et al. have tried to
constrain the deceleration parameter by using the simple ansatz
$q(z) = q_0 + (dq/dz)_{z = 0} z$ or resorting to a fourth order
expansion of the scale factor thus estimating also the jerk and
snap parameters \cite{Visser}. While the (quite large) constraints
on $q_0$ shown in their Fig.\,6 agree with our own in Eq.(\ref{eq:
qzrpl}), the vanishing of $(dq/dz)_{z = 0}$ is clearly ruled out.
It is interesting to notice, however, that a similar analysis
performed in Ref.\,\cite{Moncy} expanding the scale factor up to
the fifth order and using no priors at all gives different
results. A glance at Fig.\,2 in that paper shows that our ranges
for $q_0$ are indeed acceptable even if the best fit value quoted
there ($q_0 = -0.76$) is outside our $2 \sigma$ interval.
Moreover, Fig.\,3 of the same paper suggests that the jerk
parameter is only weakly constrained and may be also consistent
with a null value so that it is not possible to reject models with
constant $q(z)$.

Actually, there is some evidence in favor of the model. First, the
estimated Hubble constant is in good agreement with recent values
quoted in literature. In the framework of the concordance model, a
combined analysis of the CMBR anisotropy spectrum measured by
WMAP, the power spectrum of SDSS galaxies, the SNeIa Gold dataset,
the dependence of the bias on luminosity and the Ly$\alpha$ power
spectrum lead Seljak et al. to finally estimate $h =
0.710^{+0.075}_{-0.067}$ (at $99\%$ CL) \cite{Sel04} consistent
with our range in Eq.(\ref{eq: hrpl}). Results in agreement with
those of Seljak et al. (but with larger uncertainties) have also
been obtained by applying the same method to less complete dataset
and are not reported here for sake of shortness (see, e.g.,
\cite{WMAP,Teg03} and references therein). It is even more
appealing the agreement among our estimate of $h$ and those coming
from model independent methods. For instance, by combining
different calibrated local distance indicatores, the HST Key
project finally furnish $h = 0.72 {\pm} 0.08$ \cite{HSTKey} in
quite good agreement with our results. This conclusion is further
strenghtened when comparing to the results from time delays in
lensed quasars \cite{H0lens} and Sunyaev\,-\,Zel'dovich effect in
galaxy clusters \cite{H0SZ}.

Having constrained with the likelihood analysis both $h$ and
$\Omega_b/\Omega_M$, we may derive $\Omega_M$ by using an
independent estimate of $\Omega_b$. Following Kirkman et al.
\cite{Kirk}, we adopt\,:

\begin{figure}
\centering \resizebox{8.5cm}{!}{\includegraphics{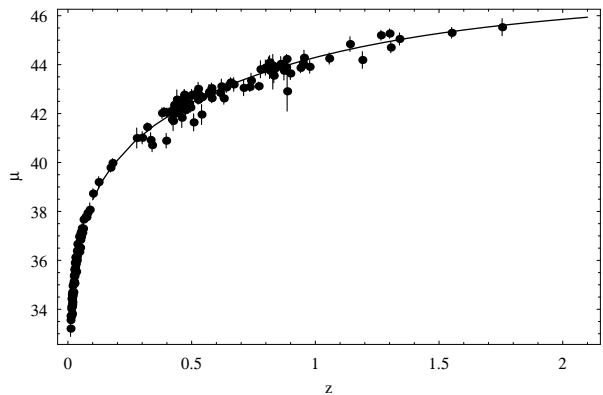}}
\caption{Best fit curve to the SNeIa Hubble diagram for the
logarithmic Lagrangian model.} \label{fig: snefitlog}
\end{figure}

\begin{displaymath}
\Omega_b h^2 = 0.0214 {\pm} 0.0020
\end{displaymath}
that, combined with Eqs.(\ref{eq: hrpl}) and Eq.(\ref{eq: obompl}), gives\,:

\begin{equation}
\Omega_M = 0.28 {\pm} 0.02
\label{eq: ompl}
\end{equation}
where the error has been roughly evaluated by propagating the $1
\sigma$ uncertainties on $h$, $\Omega_b/\Omega_M$ and $\Omega_b
h^2$ approximated as symmetric around the best fit
values\footnote{It is likely that this method underestimates the
true error thus only giving an order of magnitude estimate.}.
Eq.(\ref{eq: ompl}) is in very good agreement with recent results.
As already quoted above, using only the SNeIa Gold dataset, Riess
et al. have found $0.29^{+0.05}_{-0.03}$ for a flat $\Lambda$CDM
model, while the analysis of Seljak et al. gives $\Omega_M =
0.284_{-0.060}^{0.079}$ (at $99\%$ CL). Finally, fitting to the
$f_{gas}$ data only with priors on both $h$ and $\Omega_B h^2$,
but not imposing the flatness condition {\it ab initio}, A04
estimates $\Omega_M = 0.245^{+0.040}_{-0.037}$, while including
the CMB data, they get $\Omega_M = 0.26^{+0.06}_{-0.04}$. All
these results are in almost perfect agreement with our estimate of
$\Omega_M$ which is indeed a remarkable success.

Finally, we could use the estimated values of $n$, $h$ and
$\Omega_M$ and the age of the universe $t_0$ to put constraints on
the coupling constant $\beta$ through Eq.(\ref{eq: tzpl}).
However, this does not give us any useful information since we
have no theoretical motivation that may suggest us what is the
value of $\beta$. On the other hand, the freedom we have in the
choice of $\beta$ leaves us open the possibility to find a $R^n$
model which fits both the SNeIa Hubble diagram and the $f_{gas}$
data and also predicts the right value of $t_0$.

\subsection{$f(R) = \alpha \log{R}$}

Let us now discuss briefly the results for models with the
logarithmic Lagrangian in Eq.(\ref{eq: logfr}). With the following
choice of the model parameters\,:

\begin{figure}
\centering
\resizebox{8.5cm}{!}{\includegraphics{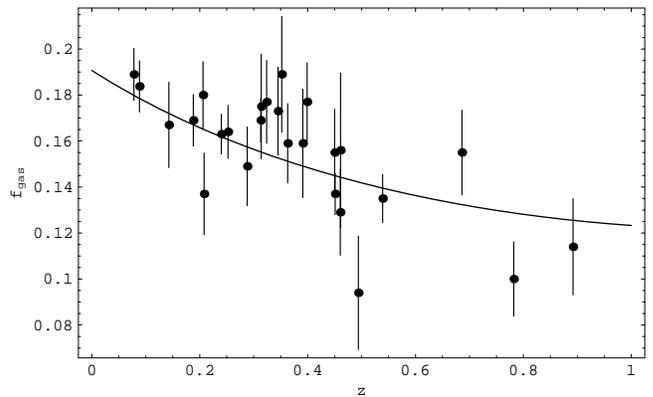}}
\caption{Best fit curve to the $f_{gas}$ data for the power law
Lagrangian model.} \label{fig: fgasfitlog}
\end{figure}

\begin{equation}
\Omega_M H_0^2 \alpha^{-1} = 0.162 \ , \ h = 0.650 \ , \ \Omega_b/\Omega_M = 0.184
\label{eq: bflog}
\end{equation}
we get the best fit curves shown in Figs.\,\ref{fig: snefitlog}
and \ref{fig: fgasfitlog}. While both fits are indeed very good,
it is interesting to note that the SNeIa Hubble diagram is now
reproduced with great accuracy also for the two SNeIa with the
highest redshift in contrast with what is observed for the power
law Lagrangian models. This is likely a consequence of having this
class of model a non constant deceleration parameter in agreement
with what is suggested by Riess et al. (within the caveat noted
above).

\begin{figure}
\centering \resizebox{8.5cm}{!}{\includegraphics{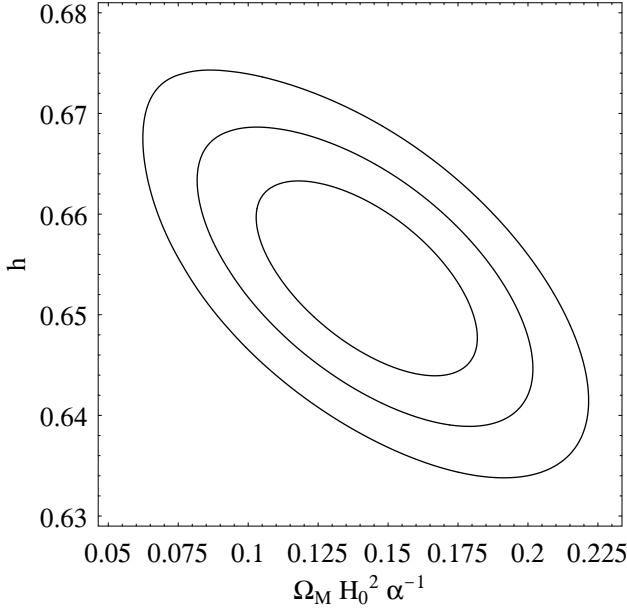}}
\caption{$1, \ 2, \ {\rm and} \ 3 \sigma$ confidence regions in the two dimensional parameter space $(\Omega_M H_0^2 \alpha^{-1}, h)$.}
\label{fig: omhztsqh}
\end{figure}

\begin{figure}
\centering \resizebox{8.5cm}{!}{\includegraphics{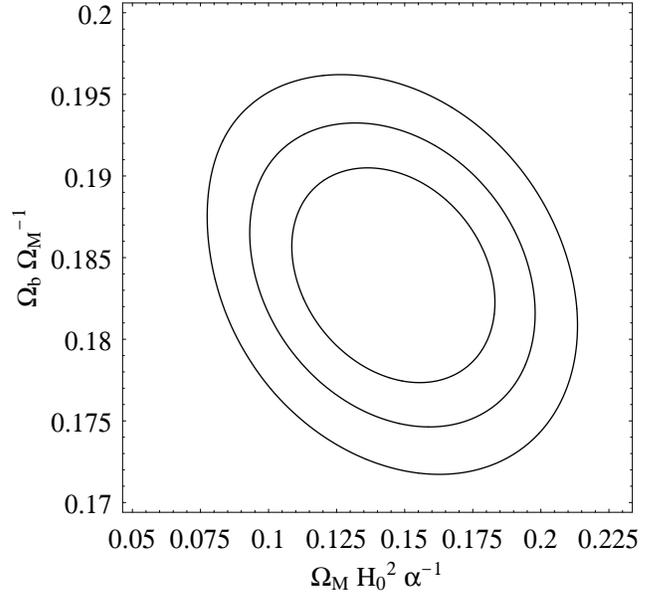}}
\caption{$1, \ 2, \ {\rm and} \ 3 \sigma$ confidence regions in the two dimensional parameter space
$(\Omega_M H_0^2 \alpha^{-1}, \Omega_b/\Omega_M)$.}
\label{fig: omhztsqobom}
\end{figure}

Figs.\,\ref{fig: omhztsqh} and \ref{fig: omhztsqobom} show the two
dimensional projections of the $1, \ 2,$ and $3 \sigma$ confidence
regions on the subset parameter space $(\Omega_M H_0^2
\alpha^{-1}, h)$ and $(\Omega_M H_0^2 \alpha^{-1},
\Omega_b/\Omega_M)$ respectively. It is clear that $\Omega_M H_0^2
\alpha^{-1}$ anticorrelates with both $h$ and $\Omega_b/\Omega_M$.
From the projection on the $(h, \Omega_b/\Omega_M)$ plane (not
shown here), we see that these parameters are negatively
correlated. Combining these plots, we may argue that the Hubble
constant is positively correlated with both $\Omega_M$ and
$\alpha$ so that the anticorrelation with $\Omega_M H_0^2
\alpha^{-1}$ is due to the degeneracy between $h$ and $\alpha$
that turns out to be stronger than that between $h$ and
$\Omega_M$.

Let us now consider the constraints on the single parameters. We get\,:

\begin{equation}
\Omega_M H_0^2 \alpha^{-1} \in \left \{
\begin{array}{ll}
(0.148, 0.174) & {\rm at} \ 1 \sigma \\ ~ & ~ ~ ~ \ \ \ \ \ \ \ ; \\
(0.129, 0.194) & {\rm at} \ 2 \sigma
\end{array}
\right .
\label{eq: oharlog}
\end{equation}

\begin{equation}
h \in \left \{
\begin{array}{ll}
(0.644, 0.657) & {\rm at} \ 1 \sigma \\ ~ & ~ ~ ~ \ \ \ \ \ \ \ ; \\
(0.637, 0.664) & {\rm at} \ 2 \sigma
\end{array}
\right .
\label{eq: hrlog}
\end{equation}

\begin{equation}
\Omega_b/\Omega_M \in \left \{
\begin{array}{ll}
(0.180, 0.188) & {\rm at} \ 1 \sigma \\ ~ & ~ ~ ~ \ \ \ \ \ \ \ . \\
(0.176, 0.192) & {\rm at} \ 2 \sigma
\end{array}
\right .
\label{eq: obomlog}
\end{equation}
It is more useful to translate the constraint on the combined
parameter $\Omega_M H_0^2 \alpha^{-1}$ (whose physical meaning is
not immediate) in a range for the present day value of the
deceleration parameter. Using Eq.(\ref{eq: qlog}), we get $q_0 =
-0.55$ as best fit value, while the confidence regions turn out to
be\,:

\begin{equation}
q_0 \in \left \{
\begin{array}{ll}
(-0.56, -0.54) & {\rm at} \ 1 \sigma \\ ~ & ~ ~ ~ \ \ \ \ \ \ \ . \\
(-0.58, -0.52) & {\rm at} \ 2 \sigma
\end{array}
\right .
\label{eq: qzrlog}
\end{equation}
Moreover, being $q(z)$ no longer constant for this class of
models, we may also estimate the transition redshift obtaining
$z_T = 0.61$ as best fit value and the following confidence
regions\,:

\begin{equation}
z_T \in \left \{
\begin{array}{ll}
(0.57, 0.66) & {\rm at} \ 1 \sigma \\ ~ & ~ ~ ~ \ \ \ \ \ \ \ . \\ (0.52,
0.74) & {\rm at} \ 2 \sigma
\end{array}
\right .
\label{eq: ztrlog}
\end{equation}
Even if the deceleration parameter is varying with the redshift
$z$, our estimate of $q_0$ is still in disagreement with the
estimates discussed in the previous subsection. As a general
remark, we notice that, as for the class of models with power law
Lagrangian, the estimated $q_0$ is higher than what is predicted
by the best fit $\Lambda$CDM model. However,  the disagreement is
now less severe and, indeed, a marginal agreement may be sometimes
obtained by considering the $3 \sigma$ confidence regions.

We may also compare the transition redshift that, for a flat
$\Lambda$CDM model, is given by\,: $z_T = [2 (1 -
\Omega_M)/\Omega_M]^{1/3} - 1$. Using, for instance, the estimate
of $\Omega_M$ given by Seljak et al., we get $z_T \in (0.52,
0.91)$ with $z_T = 0.71$ as best fit in quite a good agreement
with our Eq.(\ref{eq: ztrlog}). Moreover, it is encouraging that
our $1 \sigma$ confidence region has a non null overlap with that
estimated by Riess et al., $z_T = 0.46 {\pm} 0.13$, using the
model independent parametrization of $q(z) = q_0 + (dq/dz)_{z=0}
z$.

Regarding the Hubble constant, the confidence regions for $h$ are
almost the same as those obtained for the power law Lagrangian
case. Hence, we are still in agreement with previous results in
literature. This is not very surprising since $h$ is essentially
determined by the fit to the low redshift SNeIa and, in this
range, both $D_L$ and $D_A$ are almost model independent. As a
consequence, the estimated $h$ turns out to be the same whatever
is the underlying cosmology and in agreement with what one should
obtain by a linear fit to the $z \le 0.1$ SNeIa data.

From the constraints (\ref{eq: hrlog}) and (\ref{eq: obomlog})
and the value of $\Omega_b h^2$ in Ref.\,\cite{Kirk}, we estimate\,:

\begin{equation}
\Omega_M = 0.27 {\pm} 0.03
\label{eq: omlog}
\end{equation}
with the error evaluated as for that in Eq.(\ref{eq: ompl}). This
is in perfect agreement with both the result for the power law
Lagrangian case and the other estimates quoted above.

One could use Eq.(\ref{eq: omlog}) and the constraints on
$\Omega_M H_0^2 \alpha^{-1}$ and $h$ to narrow the range for the
coupling parameter $\alpha$. However, this does not give any
useful information since there is no way to theoretically predict
the value of $\alpha$. It is, on the contrary, more interesting to
evaluate the present age of the universe using Eq.(\ref{eq:
tzlog}) and the constraints in Eqs.(\ref{eq: oharlog}) and
(\ref{eq: hrlog}). For the best fit model, it is $t_0 = 10.3 \
{\rm Gyr}$, while $t_0$ ranges between $10 \ {\rm and} \ 11 \ {\rm
Gyr}$ for the parameters running in their $1 \sigma$ confidence
regions. These values are too low when compared to the estimated
$t_0$ for the best fit $\Lambda$CDM model. For instance, the best
fit vanilla model of Tegmark et al. \cite{Teg03} predicts $t_0 =
13.54_{-0.27}^{+0.23} \ {\rm Gyr}$ more than $9 \sigma$ higher
than our estimated upper value. Notice, however, that the
disagreement is less severe (but still of high significance) if
compared to $t_0 = 14.4_{-1.3}^{+1.4} \ {\rm Gyr}$ determined by
Rebolo et al. \cite{VSA} by fitting the $\Lambda$CDM model to the
anisotropy spectrum measured by WMAP and VSA and to the clustering
properties of 2dFGRS galaxies. However, even if in agreement with
those obtained by completely different methods, these estimates
are model dependent. Actually, our predicted $t_0$ is not
unreasonably low if we consider that globular clusters data lead
to $t_0 = 12.6_{-2.6}^{+3.4} \ {\rm Gyr}$ \cite{Krauss}, while a
lower limit $t_0 > 12.5 {\pm} 3.5 \ {\rm Gyr}$ is obtained by
nucleochronology \cite{Cayrel}. Considering the $2 \sigma$
confidence regions for the parameters $\Omega_M H_0^2 \alpha^{-1}$
and $h$, it is therefore possible to find models that are able to
successfully fit the astrophysical data we are considering (even
if they are not the preferred ones) and also predict a present age
of the universe that is not in disagreement with cosmology
independent estimates of $t_0$.

\section{Discussion and Conclusions}

Assuming that the Palatini (first order) approach is the correct
way to treat $f(R)$ theories, we have investigated the viability
of two different class of cosmological models corresponding to two
popular choices of $f(R)$, namely a power law in the scalar
curvature and a logarithmic function of $R$. The expansion rate $H
= \dot{a}/a$ may be analytically expressed as a function of the
redshift $z$ for both classes of models so that it is possible to
contrast the model predictions against the observations. In
particular, we have used the SNeIa Hubble diagram and the data on
the gas mass fraction in relaxed galaxy clusters to investigate
the viability of each class as dark energy alternative and to
constrain their parameters. The main results are sketched below.

\begin{enumerate}

\item{Both classes of models provide very good fits to the data
even if the choice $f(R) = \alpha \ln{R}$ leads to a Hubble
diagram that is in better agreement with the trend shown by the
highest redshift SNeIa. However, the paucity of the data does not
allow us to eventually prefer one model to the other.}

\item{Eqs.(\ref{eq: bfpl}) and (\ref{eq: bflog}) give the best
fit parameters for the power law and logarithmic Lagrangian models
respectively. The confidence regions have been determined from the
marginalized likelihoods and are reported in Eqs.(\ref{eq:
nrpl})\,-\,(\ref{eq: obompl}) for the models with $f(R) = \beta
R^n$ and in Eqs.(\ref{eq: oharlog})\,-\,(\ref{eq: obomlog}) for
those with $f(R) = \alpha \ln{R}$. To better compare the model
predictions with previous results in literature, we have evaluated
the present day deceleration parameter $q_0$ and the matter
density parameter $\Omega_M$ (assuming the estimate of $\Omega_b
h^2$ in Ref.\,\cite{Kirk}). For both classes of models, $q_0$
turns out to be higher than what is predicted for the concordance
$\Lambda$CDM model, i.e. $f(R)$ theories lead to models that
accelerate less than what is observed. This result is however
somewhat weakened by comparing with model independent estimates of
$q_0$ even if these latter may be affected by systematic errors.
As far as the matter content is concerned, for both classes of
models $\Omega_M$ is in very good agreement with what is inferred
from galaxy clusters and estimated by fitting the $\Lambda$CDM
model to the available astrophysical data.}

\item{To ameliorate the agreement with the observed $q_0$,
one should increase the value of $n$ for the models with power law
Lagrangians or decrease that of $\Omega_M H_0^2 \alpha^{-1}$ for
models with $f(R) = \alpha \ln{R}$. In this case, a good fit to the data
may still be obtained provided that both $h$ and $\Omega_b/\Omega_M$ are increased.
While higher values of $h$ could still be compatible with the local estimates of the
Hubble constant, increasing $\Omega_b/\Omega_M$ leads to lower values of $\Omega_M$.
 Actually, the very good agreement among the estimated $\Omega_M$ in Eqs.(\ref{eq: ompl})
  and (\ref{eq: omlog}) and the results in literature is a strong evidence against this
  choice. Therefore, we conclude that it is not possible to recover the same value of $q_0$
  in the concordance model by using power law or logarithmic Lagrangians.}

\item{A model independent estimate of the present day age of the universe $t_0$ allows one
to break the matter/geometry degeneracy inherent in $f(R)$
theories recovering the value of the coupling constant. For power
law Lagrangians, this is indeed the only way to determine $\beta$
thus offering the possibility to always recover a model that both
fits the SNeIa Hubble diagram and the data on the gas mass
fraction in relaxed galaxy clusters and also has the correct age.
On the other hand, $t_0$ is an independent check for models with
logarithmic Lagrangian since, in this case, it may be evaluated as
a function of the two parameters $\Omega_M H_0^2 \alpha^{-1}$ and
$h$ and compared with previous results in literature. It turns out
that the predicted $t_0$ is lower than the value estimated for the
$\Lambda$CDM model and only marginally consistent with what is
inferred from globular clusters and nucleochronology.}

\end{enumerate}

The results summarized above may pave the way to the solution of
an intriguing dilemma\,: is Einsteinian general relativity the
correct theory of gravity? If yes, then dark energy is absolutely
needed to explain the accelerated expansion of the universe and
hence all the theoretical efforts of cosmologists have to be
dedicated to understanding its nature. On the contrary, if $f(R)$
theories are indeed able to explain the accelerated expansion,
then it is time to investigate in more detail what is the right
choice for the function $f(R)$ and how the variation has to be
performed (higher order metric or first order Palatini approach).

From the observational point of view we have adopted here, there
are no strong evidences against models with power law or
logarithmic Lagrangians in the framework of the Palatini approach.
On the contrary, we have seen that both classes of models
successfully fit the data with values of the Hubble constant and
matter content in good agreement with some model independent
estimates. However, there are some hints that could lead to reject
both choices for $f(R)$. Models with power law Lagrangians have a
constant $q(z)$ so that they are always accelerating. This is not
consistent with the (tentatively) observed transition from
acceleration to deceleration at $z_T \simeq 0.5$. Moreover, a
constant $q(z)$ could give rise to problems with nucleosynthesis
and structure formation. On the contrary, models with a
logarithmic Lagrangian are not affected by such problems and
indeed they predicts a transition redshift which is in good
agreement with the estimates for the $\Lambda$CDM model. On the
other hand, these models turn out to be too young, i.e. $t_0$ is
lower than what is expected.

Actually, a more general remark is in oder here. Let us suppose we
have found that a given choice for $f(R)$ leads to models that are
in agreement with the data so that we should conclude that this
class of models correctly describe the present day universe. What
about the early universe? One could expect that the functional
expression of $f(R)$ is not changing during the evolution of the
universe, even if $R$ may evolve with cosmic time. If this were
the case, then the correct choice for $f(R)$ should be the one
that leads to models that are not only able to reproduce the
phenomenology we observe today, but also give rise to an
inflationary period in the early universe. Therefore, we should
reject logarithmic Lagrangians since it is well known they do not
predict any inflationary period. On the other hand, the choice
$f(R) = \beta R^n$ is able to explain inflation provided one sets
$n = 2$, not too far from our estimate in Eq.(\ref{eq: nrpl}).
From this point of view, it is worth noticing that the
astrophysical data we have considered probe only the present day
universe, while $t_0$ depends on the full evolutionary history.
Indeed, logarithmic Lagrangians fail to reproduce the correct
$t_0$ in the same way as they fail to give rise to inflation,
while both inflation and $t_0$ are correctly predicted by models
with power law $f(R)$. This may argue in favour of this choice for
$f(R)$, but actually there is no reason to exclude the possibility
that also the functional expression of $f(R)$ changes with time so
that neither class of models may be definitively rejected or
deemed as the correct one from this point of view.

Summarizing, the likelihood analysis presented here allows us to
conclude that the Palatini approach to $f(R)$ theories leads to
models that are able to reproduce both the SNeIa Hubble diagram
and the data on the gas mass fraction in galaxy clusters. From an
observational point of view, this means that both power law and
logarithmic $f(R)$ are viable candidates to explain the observed
accelerated expansion without the need of any kind of dark energy.
However, open questions are still on the ground.

First, we have not yet been able to discriminate between the two
classes of models. Theoretical considerations and some hints from
the age of the universe could argue in favour of the power law
$f(R)$, while the observed transition from acceleration to
deceleration in the past disfavors this choice. To solve this
issue, one has to resort to high redshift probes such as the CMBR
anisotropy spectrum. While the data are of superb quality, the
underlying theory is still to be developed so that fitting the
CMBR anisotropy temperature and polarization spectra with $f(R)$
theories will be  quite a demanding task.

Second, we have only considered two physically motivated and
popular choices for $f(R)$. Several other models are possible and
are worth of being tested against the data. In particular the
$R\ln R$ Lagrangian which is related to the Straobinsky
inflationary model \cite{starobinsky} and to the limit
$R^n\rightarrow R$ for $n\simeq 1$ being \cite{CF}
\begin{equation}
R^{1+\epsilon}=RR^{\epsilon}=R\left(e^{\epsilon\ln R}\right)\simeq
R+\epsilon R\ln R +{\cal O}(\epsilon^2)\,.\end{equation}

However, rather than being confused by a plethora of successfull
models, it is desiderable to develop a method that allows to
directly reconstruct $f(R)$ from the data with as less as possible
aprioristic assumptions. This will be the subject of a forthcoming
paper \cite{CCF}.

Last but not least, whether the Palatini approach is indeed the
correct method to treat $f(R)$ theories or the metric approach
should be preferred is still an unsolved problem. We have shown
here that the Palatini approach is not rejected by the data, but a
similar analysis for the same models considered in the framework
of the metric approach is still lacking. However, it is worth
noticing that even this test will not be conclusive. Let us
consider, for instance, two choices $f_1(R)$ and $f_2(R)$ and let
us suppose that $f_1(R)$ fit the data if considered in the
framework of the metric approach, but not if the Palatini approach
is used. Let us further assume that the opposite holds for
$f_2(R)$. From an observational point of view, it is impossible to
select between $f_1(R)$ and $f_2(R)$. Hence, observations could
never suggest what is the correct way of performing the variation
of a $f(R)$ Lagrangian. The answer to this question is outside the
possibilities of an astronomer and lies fully in the field of a
theoretician.

As a final comment, we would like to stress the need for synergy
between theory and observations. While it is possible to build a
physically motivated and mathematically elegant theory, it is not
so easy to fit the significant amount of astrophysical data now
available. Since the words {\it observational} and {\it cosmology}
may today be joined together in a single meaningful term ({\it
observational cosmology}), it is time to look at every
theoretician's proposal from an observational point of view before
drawing any conclusion about the validity of a whatever model.
Even if not always conclusive, in our opinion, this is the only
way to shed light on the dark side of the universe.

\acknowledgments{We warmly thank R.W. Schmidt for having given us
in electronic form the data on the gas mass fraction in advance of
publication. We also acknowledge G. Allemandi, A. Borowiec, M.
Capone and A. Troisi for  interesting discussions on the topic.}

\end{document}